\documentstyle[10pt,epsf,epsfig,dp_delphititle,rotating,amsmath]
{dp_delphi}
\makeindex
\def\DpPaperGroup{PH-EP}
\def\DpPaperRef{2004-18}
\def\DpDate{3 May 2004}
\def\DpAuthors{DELPHI Collaboration}
\def\DpSubmit{(Accepted by Physics Letters B)}
\def\DpTitle{{
Coherent Soft Particle Production \\
in $\mathbf Z $ Decays into Three Jets
   }}
\def\DpComment{~}\def\DpEMail{~}
\newcommand{\gev}{\mbox{\section{Motivation and Theoretical Basis\\ of the Measurement}\,Ge\kern-0.2exV}}
\newcommand{\mev}{\mbox{\,Me\kern-0.2exV}}

\newcommand{\epem}{${e^+e^-}$}
\newcommand{\cacf}{$C_A/C_F$}

\newcommand{\beq}{\begin{equation}}
\newcommand{\eeq}{\end{equation}}
\newcommand{\degs}{$^{\circ}$}
\def\jetset{{\sc Jetset}}

\def\delphi{{\sc Delphi}}
\def\lep{{\sc Lep}}
\def\cern{{\sc CERN}}

\begin{document}
\makeatletter
\newcount\@tempcntc
\def\@citex[#1]#2{\if@filesw\immediate\write\@auxout{\string\citation{#2}}\fi
  \@tempcnta\z@\@tempcntb\m@ne\def\@citea{}\@cite{\@for\@citeb:=#2\do
    {\@ifundefined
       {b@\@citeb}{\@citeo\@tempcntb\m@ne\@citea\def\@citea{,}{\bf ?}\@warning
       {Citation `\@citeb' on page \thepage \space undefined}}%
    {\setbox\z@\hbox{\global\@tempcntc0\csname b@\@citeb\endcsname\relax}%
     \ifnum\@tempcntc=\z@ \@citeo\@tempcntb\m@ne
       \@citea\def\@citea{,}\hbox{\csname b@\@citeb\endcsname}%
     \else
      \advance\@tempcntb\@ne
      \ifnum\@tempcntb=\@tempcntc
      \else\advance\@tempcntb\m@ne\@citeo
      \@tempcnta\@tempcntc\@tempcntb\@tempcntc\fi\fi}}\@citeo}{#1}}
\def\@citeo{\ifnum\@tempcnta>\@tempcntb\else\@citea\def\@citea{,}%
  \ifnum\@tempcnta=\@tempcntb\the\@tempcnta\else
   {\advance\@tempcnta\@ne\ifnum\@tempcnta=\@tempcntb \else \def\@citea{--}\fi
    \advance\@tempcnta\m@ne\the\@tempcnta\@citea\the\@tempcntb}\fi\fi}
 
\makeatother
\begin{titlepage}
\pagenumbering{roman}
\CERNpreprint{\DpPaperGroup}{\DpPaperRef} 
\date{{\small\DpDate}} 
\title{\DpTitle} 
\address{\DpAuthors} 
\begin{shortabs} 
\noindent
\noindent 
Low-energy particle production perpendicular to the event plane 
in three-jet events produced in $Z$ decays in \epem\ annihilation 
is measured and compared to that perpendicular to the event axis in 
two-jet events. The topology 
dependence of the hadron production ratio is found to
agree with a leading-order QCD prediction. This agreement and 
especially the need for the presence of a destructive interference term
gives evidence for the coherent nature of gluon radiation.
Hadron production in three-jet events is found to be
directly proportional to a single topological scale
function of the inter-jet angles. The slope of the dependence of 
the multiplicity
with respect to the topological scale 
was measured to be:
\begin{equation*}
2.211 \pm 0.014\text{(stat.)} \pm 0.053\text{(syst.)} ~~~
\end{equation*}
in 
good agreement with the expectation given by the 
colour-factor ratio $C_A/C_F=9/4$.
This result strongly supports the assumption of local parton-hadron duality,
LPHD, at low hadron momentum. 

\end{shortabs}
\vfill
\begin{center}
\DpSubmit \ \\ 
\DpComment \ \\
\DpEMail \ \\
\end{center}
\vfill
\clearpage
\headsep 10.0pt
\addtolength{\textheight}{10mm}
\addtolength{\footskip}{-5mm}
\begingroup
%
\newcommand{\DpName}[2]{\hbox{#1$^{\ref{#2}}$},\hfill}
\newcommand{\DpNameTwo}[3]{\hbox{#1$^{\ref{#2},\ref{#3}}$},\hfill}
\newcommand{\DpNameThree}[4]{\hbox{#1$^{\ref{#2},\ref{#3},\ref{#4}}$},\hfill}
\newskip\Bigfill \Bigfill = 0pt plus 1000fill
\newcommand{\DpNameLast}[2]{\hbox{#1$^{\ref{#2}}$}\hspace{\Bigfill}}
\small
\noindent
\DpName{J.Abdallah}{LPNHE}
\DpName{P.Abreu}{LIP}
\DpName{W.Adam}{VIENNA}
\DpName{P.Adzic}{DEMOKRITOS}
\DpName{T.Albrecht}{KARLSRUHE}
\DpName{T.Alderweireld}{AIM}
\DpName{R.Alemany-Fernandez}{CERN}
\DpName{T.Allmendinger}{KARLSRUHE}
\DpName{P.P.Allport}{LIVERPOOL}
\DpName{U.Amaldi}{MILANO2}
% ...
\normalsize
\endgroup
\titlefoot{Department of Physics and Astronomy, Iowa State
     University, Ames IA 50011-3160, USA
    \label{AMES}}
\titlefoot{Physics Department, Universiteit Antwerpen,
     Universiteitsplein 1, B-2610 Antwerpen, Belgium \\
     \indent~~and IIHE, ULB-VUB,
     Pleinlaan 2, B-1050 Brussels, Belgium \\
     \indent~~and Facult\'e des Sciences,
     Univ. de l'Etat Mons, Av. Maistriau 19, B-7000 Mons, Belgium
    \label{AIM}}
\titlefoot{Physics Laboratory, University of Athens, Solonos Str.
     104, GR-10680 Athens, Greece
    \label{ATHENS}}
\titlefoot{Department of Physics, University of Bergen,
     All\'egaten 55, NO-5007 Bergen, Norway
    \label{BERGEN}}
\titlefoot{Dipartimento di Fisica, Universit\`a di Bologna and INFN,
     Via Irnerio 46, IT-40126 Bologna, Italy
    \label{BOLOGNA}}
\titlefoot{Centro Brasileiro de Pesquisas F\'{\i}sicas, rua Xavier Sigaud 150,
     BR-22290 Rio de Janeiro, Brazil \\
     \indent~~and Depto. de F\'{\i}sica, Pont. Univ. Cat\'olica,
     C.P. 38071 BR-22453 Rio de Janeiro, Brazil \\
     \indent~~and Inst. de F\'{\i}sica, Univ. Estadual do Rio de Janeiro,
     rua S\~{a}o Francisco Xavier 524, Rio de Janeiro, Brazil
    \label{BRASIL}}
\titlefoot{Coll\`ege de France, Lab. de Physique Corpusculaire, IN2P3-CNRS,
     FR-75231 Paris Cedex 05, France
    \label{CDF}}
\titlefoot{CERN, CH-1211 Geneva 23, Switzerland
    \label{CERN}}
\titlefoot{Institut de Recherches Subatomiques, IN2P3 - CNRS/ULP - BP20,
     FR-67037 Strasbourg Cedex, France
    \label{CRN}}
\titlefoot{Now at DESY-Zeuthen, Platanenallee 6, D-15735 Zeuthen, Germany
    \label{DESY}}
\titlefoot{Institute of Nuclear Physics, N.C.S.R. Demokritos,
     P.O. Box 60228, GR-15310 Athens, Greece
    \label{DEMOKRITOS}}
\titlefoot{FZU, Inst. of Phys. of the C.A.S. High Energy Physics Division,
     Na Slovance 2, CZ-180 40, Praha 8, Czech Republic
    \label{FZU}}
\titlefoot{Dipartimento di Fisica, Universit\`a di Genova and INFN,
     Via Dodecaneso 33, IT-16146 Genova, Italy
    \label{GENOVA}}
\titlefoot{Institut des Sciences Nucl\'eaires, IN2P3-CNRS, Universit\'e
     de Grenoble 1, FR-38026 Grenoble Cedex, France
    \label{GRENOBLE}}
\titlefoot{Helsinki Institute of Physics, P.O. Box 64,
     FIN-00014 University of Helsinki, Finland
    \label{HELSINKI}}
\titlefoot{Joint Institute for Nuclear Research, Dubna, Head Post
     Office, P.O. Box 79, RU-101 000 Moscow, Russian Federation
    \label{JINR}}
\titlefoot{Institut f\"ur Experimentelle Kernphysik,
     Universit\"at Karlsruhe, Postfach 6980, DE-76128 Karlsruhe,
     Germany
    \label{KARLSRUHE}}
\titlefoot{Institute of Nuclear Physics,Ul. Kawiory 26a,
     PL-30055 Krakow, Poland
    \label{KRAKOW1}}
\titlefoot{Faculty of Physics and Nuclear Techniques, University of Mining
     and Metallurgy, PL-30055 Krakow, Poland
    \label{KRAKOW2}}
\titlefoot{Universit\'e de Paris-Sud, Lab. de l'Acc\'el\'erateur
     Lin\'eaire, IN2P3-CNRS, B\^{a}t. 200, FR-91405 Orsay Cedex, France
    \label{LAL}}
\titlefoot{School of Physics and Chemistry, University of Lancaster,
     Lancaster LA1 4YB, UK
    \label{LANCASTER}}
\titlefoot{LIP, IST, FCUL - Av. Elias Garcia, 14-$1^{o}$,
     PT-1000 Lisboa Codex, Portugal
    \label{LIP}}
\titlefoot{Department of Physics, University of Liverpool, P.O.
     Box 147, Liverpool L69 3BX, UK
    \label{LIVERPOOL}}
\titlefoot{Dept. of Physics and Astronomy, Kelvin Building,
     University of Glasgow, Glasgow G12 8QQ
    \label{GLASGOW}}
\titlefoot{LPNHE, IN2P3-CNRS, Univ.~Paris VI et VII, Tour 33 (RdC),
     4 place Jussieu, FR-75252 Paris Cedex 05, France
    \label{LPNHE}}
\titlefoot{Department of Physics, University of Lund,
     S\"olvegatan 14, SE-223 63 Lund, Sweden
    \label{LUND}}
\titlefoot{Universit\'e Claude Bernard de Lyon, IPNL, IN2P3-CNRS,
     FR-69622 Villeurbanne Cedex, France
    \label{LYON}}
\titlefoot{Dipartimento di Fisica, Universit\`a di Milano and INFN-MILANO,
     Via Celoria 16, IT-20133 Milan, Italy
    \label{MILANO}}
\titlefoot{Dipartimento di Fisica, Univ. di Milano-Bicocca and
     INFN-MILANO, Piazza della Scienza 2, IT-20126 Milan, Italy
    \label{MILANO2}}
\titlefoot{IPNP of MFF, Charles Univ., Areal MFF,
     V Holesovickach 2, CZ-180 00, Praha 8, Czech Republic
    \label{NC}}
\titlefoot{NIKHEF, Postbus 41882, NL-1009 DB
     Amsterdam, The Netherlands
    \label{NIKHEF}}
\titlefoot{National Technical University, Physics Department,
     Zografou Campus, GR-15773 Athens, Greece
    \label{NTU-ATHENS}}
\titlefoot{Physics Department, University of Oslo, Blindern,
     NO-0316 Oslo, Norway
    \label{OSLO}}
\titlefoot{Dpto. Fisica, Univ. Oviedo, Avda. Calvo Sotelo
     s/n, ES-33007 Oviedo, Spain
    \label{OVIEDO}}
\titlefoot{Department of Physics, University of Oxford,
     Keble Road, Oxford OX1 3RH, UK
    \label{OXFORD}}
\titlefoot{Dipartimento di Fisica, Universit\`a di Padova and
     INFN, Via Marzolo 8, IT-35131 Padua, Italy
    \label{PADOVA}}
\titlefoot{Rutherford Appleton Laboratory, Chilton, Didcot
     OX11 OQX, UK
    \label{RAL}}
\titlefoot{Dipartimento di Fisica, Universit\`a di Roma II and
     INFN, Tor Vergata, IT-00173 Rome, Italy
    \label{ROMA2}}
\titlefoot{Dipartimento di Fisica, Universit\`a di Roma III and
     INFN, Via della Vasca Navale 84, IT-00146 Rome, Italy
    \label{ROMA3}}
\titlefoot{DAPNIA/Service de Physique des Particules,
     CEA-Saclay, FR-91191 Gif-sur-Yvette Cedex, France
    \label{SACLAY}}
\titlefoot{Instituto de Fisica de Cantabria (CSIC-UC), Avda.
     los Castros s/n, ES-39006 Santander, Spain
    \label{SANTANDER}}
\titlefoot{Inst. for High Energy Physics, Serpukov
     P.O. Box 35, Protvino, (Moscow Region), Russian Federation
    \label{SERPUKHOV}}
\titlefoot{J. Stefan Institute, Jamova 39, SI-1000 Ljubljana, Slovenia
     and Laboratory for Astroparticle Physics,\\
     \indent~~Nova Gorica Polytechnic, Kostanjeviska 16a, SI-5000 Nova Gorica, Slovenia, \\
     \indent~~and Department of Physics, University of Ljubljana,
     SI-1000 Ljubljana, Slovenia
    \label{SLOVENIJA}}
\titlefoot{Fysikum, Stockholm University,
     Box 6730, SE-113 85 Stockholm, Sweden
    \label{STOCKHOLM}}
\titlefoot{Dipartimento di Fisica Sperimentale, Universit\`a di
     Torino and INFN, Via P. Giuria 1, IT-10125 Turin, Italy
    \label{TORINO}}
\titlefoot{INFN,Sezione di Torino, and Dipartimento di Fisica Teorica,
     Universit\`a di Torino, Via P. Giuria 1,\\
     \indent~~IT-10125 Turin, Italy
    \label{TORINOTH}}
\titlefoot{Dipartimento di Fisica, Universit\`a di Trieste and
     INFN, Via A. Valerio 2, IT-34127 Trieste, Italy \\
     \indent~~and Istituto di Fisica, Universit\`a di Udine,
     IT-33100 Udine, Italy
    \label{TU}}
\titlefoot{Univ. Federal do Rio de Janeiro, C.P. 68528
     Cidade Univ., Ilha do Fund\~ao
     BR-21945-970 Rio de Janeiro, Brazil
    \label{UFRJ}}
\titlefoot{Department of Radiation Sciences, University of
     Uppsala, P.O. Box 535, SE-751 21 Uppsala, Sweden
    \label{UPPSALA}}
\titlefoot{IFIC, Valencia-CSIC, and D.F.A.M.N., U. de Valencia,
     Avda. Dr. Moliner 50, ES-46100 Burjassot (Valencia), Spain
    \label{VALENCIA}}
\titlefoot{Institut f\"ur Hochenergiephysik, \"Osterr. Akad.
     d. Wissensch., Nikolsdorfergasse 18, AT-1050 Vienna, Austria
    \label{VIENNA}}
\titlefoot{Inst. Nuclear Studies and University of Warsaw, Ul.
     Hoza 69, PL-00681 Warsaw, Poland
    \label{WARSZAWA}}
\titlefoot{Fachbereich Physik, University of Wuppertal, Postfach
     100 127, DE-42097 Wuppertal, Germany
    \label{WUPPERTAL}}
\addtolength{\textheight}{-10mm}
\addtolength{\footskip}{5mm}
\clearpage
\headsep 30.0pt
\end{titlepage}
%
\pagenumbering{arabic} 
\setcounter{footnote}{0} %
\large
\section{Introduction}
Quantum-mechanical interference effects are basic to the gauge theories of
fundamental interactions such as Quantum Chromodynamics, QCD.
The probabilistic Parton Model of strong interactions, 
however, neglecting interference
effects still successfully describes many processes 
as an incoherent sum of the individual sub-processes.
In fact it has proved difficult to show strong-interaction interference effects
in high-energy inelastic processes like \epem\ annihilation,
deep-inelastic scattering or $p\overline{p}$ interactions.
Evidence for coherent soft gluon emission in multiparticle production
comes from the necessity to include
angular ordering in the fragmentation models in order to describe the
energy dependence of hard interactions, from the so called {\it hump backed 
plateau} in the logarithmic scaled-momentum spectrum of hadrons
due to the suppression of low-energy particle production, and from
the string-effect in three-jet events in \epem\ annihilation explained by
destructive interference. For a comprehensive review see
\cite{khozeochsrev}.

Arguments against the conclusiveness of these verifications 
of coherence effects have been
raised, however. 
Incoherent fragmentation models involving a large number of parameters
allow for a sufficient description of the data at least at fixed
centre-of-mass energy \cite{tune-incoh}. The hadronic momentum spectrum and 
the energy dependence of the peak of this distribution may also
be accounted for when assuming a non-minimal phase-space structure
\cite{pavelvolo}. The
string-effect measurements, except for symmetric
topologies \cite{glupaper1,khozeochsrev}, are also influenced by boost effects.

In this letter evidence for quantum-mechanical interference or colour coherence
is presented from the comparison of low-energy hadron production perpendicular 
to the event plane of three-jet events and the event axis of two-jet events,
respectively.
Similarly to the string-effect, in this case the presence of destructive 
interference is expected. Presuming colour coherence, the measurement is 
also directly 
sensitive to the colour of the underlying partons, $q$, $\overline{q}$ and
$g$. 
The ratio of the
colour charges or the colour-factors for gluons and quarks, \cacf, can
be verified directly from the dependence of the hadron production on the 
three-jet topology.

This letter is organised as follows. In Section \ref{theo} a basic
description of the theoretical formulae and the ideas underlying this
measurement are given.
Section \ref{data} gives an overview of the data, selections 
and corrections applied
and describes the measurement. In Section \ref{results} results 
on the hadron multiplicity observed in cones oriented perpendicular to
the event plane of three-jet events and the event axis of two-jet events,
respectively, are presented and conclusions are given.

\section{Motivation and Theoretical Basis\\ of the Measurement\label{theo}}
The hadronisation process is believed to commence with the radiation of soft
gluons from the initial hard partons. 
Due to the higher colour charge of gluons compared to quarks, 
soft gluon radiation and consequently hadron production 
is expected to be increased in the high energy limit by the 
gluon to quark colour-factor ratio \cacf$=9/4$. 
This simple expectation is perturbed for hadrons sharing a
large fraction of the energy of the underlying parton. 
Leading-particle effects 
become evident and, as such particles are created last in time
\cite{khozemuellerdok}, effects due to
energy conservation are important. 
The latter influence is stronger for gluon jets due to the more copious
particle production.

When a radiated hard gluon stays close to the radiating quark,
additional soft gluons may be unable to resolve the individual colour
charges of the close-by partons. In this limit only coherent radiation off
the parton ensemble, that is from the colour charge of the initial quark
occurs. This fundamental effect also leads to the angular-ordering property of
soft gluon emission.

Coherence effects should be best visible for low-energy hadrons
emitted perpendicularly to the radiating partons as
these particles cannot be assigned to a specific jet and have a poor 
resolution power for close-by colour charges.

The cross-section for soft-gluon radiation in ($q\overline{q}g$)
three-jet events
has been calculated in \cite{DokKhoz}.
It varies strongly as a function of the three-jet topology.
In contrast to many other event characteristics, the soft radiation
out of  the $q\overline{q}g$ plane is dominated in the perturbative
calculation by the leading-order contribution which
 takes a particularly simple form \cite{ochskhoze}:
\begin{xalignat}{3}
\label{e:basic}
d\sigma_3 &= d\sigma_2 \cdot \frac{C_A}{C_F} \cdot r_t & 
r_t &= \frac{1}{4}\left \{ \widehat{~q~g~} + \widehat{~\overline{q}~g~} 
- \frac{1}{N_c^2}\widehat{~q~\overline{q}~} \right \} 
\intertext{with the so called {\it antenna function}} \nonumber
\widehat{~i~j~} &= 1-\cos{ \theta_{ij}} = 2 \sin^2{\frac{\theta_{ij}}{2}} ~~~~.
\end{xalignat}
Here $d\sigma_3$ is the cross-section for soft gluon emission perpendicular to
the $q\overline{q}g$ plane, $d\sigma_2$ the corresponding cross-section for
emission perpendicular to the  two-jet event ($q\overline{q}$) axis, and
$\theta_{ij}$ denotes the angle between the partons $i$ and $j$.
The complete topology dependence can be factorised in the topological
term $r_t$ which, multiplied by the energy,
obviously plays the r\^ole of the 
evolution  scale of the low energy hadron multiplicity.

\mbox{Equation \ref{e:basic}} interpolates between the two cases 
in which $q\overline{q}g$ events appear as two-jet events.
For gluon radiation parallel to the quark (or antiquark) 
$r_t=4/9$
and $d\sigma_3=d\sigma_2$, as soft particles cannot resolve the individual
colour charges of the  quark and the gluon.
In the case where the
gluon recoils with respect to the quark antiquark pair  $r_t=1$.
This case is similar to an initial two-gluon state forming a 
colour singlet and $d\sigma_3={C_A}/{C_F}\cdot d\sigma_2$.
The term inversely proportional to the square of the number of
colours, $N_c=3$, in $r_t$ is due to destructive interference and 
provides an additional suppression of particle production on top
of the boost effects caused by the string-like topology in general
three-jet events.

The interest of the measurement presented in this paper is:
\begin{itemize}
\item
the experimental test of the predicted topology dependence 
manifesting the coherent nature of hadron production,
\item
the verification of the $1/N_c^2$ term as direct evidence for the negative
interference term beyond the probabilistic parton cascade and
\item
the measurement of the slope of the homogenous straight-line given by Equation
\ref{e:basic}
and thereby the verification of the colour-factor ratio \cacf.
\end{itemize}

In order to verify   the above predictions experimentally,  charged-particle
production has been studied in cones with $\vartheta_{\text{cone}}=30^{\circ}$
half opening angle situated 
perpendicular to both sides of the three-jet plane or the two-jet event axis.
For two-jet events the azimuthal orientation of the cones is taken randomly.
In order to distinguish events with two, three and four or more jets  
the angular ordered Durham (aoD) algorithm 
is used with a jet-resolution parameter $y_{cut}=0.015$.
The aoD algorithm corresponds to the Cambridge algorithm 
without soft freezing \cite{camjet}.
It has been chosen as it is expected to lead to optimal
reconstruction of jet directions but to avoid complications due to
soft freezing.
The choice of a fixed jet-resolution is necessary as the leading-order
prediction of Equation \ref{e:basic}, applies to two- and three-jet events. 
Consequently, events with less than two or more than three jets are 
explicitely excluded from the analysis.

These experimental choices, made in order to approximate the theoretical
prediction, imply systematic uncertainties in the data-to-theory comparison.
The possible systematic variations of the data were assessed by varying
the cone opening angle in the range 
$20^{\circ} \leq \vartheta_{\text{cone}} \leq 40^{\circ}$
and the jet-resolution in the range $0.01 \leq y_{cut} \leq 0.02$.
Moreover the Cambridge \cite{camjet} and Durham \cite{durham} algorithms 
were used alternatively. 
The uncertainty of the leading-order prediction, Equation \ref{e:basic},
is not specified in \cite{ochskhoze} and, consequently, is not included 
in the uncertainties of the derived quantities quoted below.

In order to compare the prediction, Equation \ref{e:basic}, 
to the data directly, the prediction
is evaluated for each of the possible assignments of the gluon to the three
jets. Finally, the three results are averaged with each assignment
being weighted according to the respective three-jet matrix element.
The topological term 
$r_t$ 
in principle varies between 
$4/9 \sim 0.44$  and $1$. In this analysis the lower limit is 
$r_t^{\text{min}}\sim 0.5$, as the cut in the jet-resolution implies a minimum angle
between the gluon and a quark. 
As gluon-jets stay unidentified and gluon radiation at low angles
dominates,
the maximum weighted average value of $r_t$
, $r_t^{\text{max}}\sim 0.71$ is
obtained for fully-symmetric three-jet events.

\section{Data and Data Analysis\label{data}}
This analysis is based on hadronic $Z$ decays collected in the years
1992 to 1995 with the \delphi\ detector at the \lep\ \epem\ collider at
\cern.~ 
The \delphi\ detector was a hermetic collider detector with a solenoidal
magnetic field, extensive tracking facilities including a micro-vertex
detector,
electromagnetic and hadronic calorimetry as well as strong particle
identification capabilities. 
The detector and its performance are described in detail elsewhere
\cite{det,perf}.

In order to select well-measured particles originating from the interaction
point, the cuts shown in Table \ref{t:tcuts} 
were applied to the measured 
tracks and electromagnetic or hadronic calorimeter clusters.
Here $p$ and $E$ denote the particle's momentum and energy, 
$\vartheta_{\text{polar}}$ denotes the polar angle with respect to the beam, 
$\epsilon_k$ is the distance of closest approach 
to the interaction point in the plane
perpendicular to ($xy$) or along ($z$) the beam, respectively, 
$L_{\text{track}}$ is the measured track length.
$E_{\text{HPC}}$ ($E_{\text{EMF}}$) denotes the 
energy of a cluster as measured with the barrel (forward) electromagnetic 
calorimeter, 
respectively, and  $E_{\text{HAC}}$ the cluster energy measured by the
hadronic calorimeter.

\begin{table} [htb]
\begin{minipage}{7.5cm}
\begin{center}
  \begin{tabular}{|c|c|}
    \hline
    variable & cut \\
    \hline \hline
    $p$ & $ \ge 0.4~\mathrm{ GeV}$  \\
      \hline
    $\vartheta_{\mathrm{ polar}}$ &  $20^\circ-160^\circ$ \\
      \hline
    $\epsilon_{xy}$ & $\le 5 ~\mathrm{ cm}$ \\
      \hline
    $\epsilon_z$ &  $\le 10 ~\mathrm{ cm}$  \\
      \hline
    $L_{\mathrm{ track}}$ & $\ge 30~\mathrm{ cm}$  \\
      \hline
    $\Delta p/p$ & $\le 100\%$ \\
    \hline
   $E_{\text{HPC}}$ & $0.5~\mathrm{GeV} - 50~\mathrm{GeV}$  \\ \hline
   $E_{\text{EMF}}$ & $0.5~\mathrm{GeV} - 50~\mathrm{GeV}$  \\ \hline
   $E_{\text{HAC}}$ & $1~\mathrm{GeV} - 50~\mathrm{GeV}$  \\ \hline
  \end{tabular}  
  \caption{\label{t:tcuts} Selection cuts applied to charged-particle tracks
and to calorimeter clusters.}
\end{center} 
\end{minipage}
\hfill
\begin{minipage}{7cm}
\begin{center}
  \begin{tabular}{|c|c|}
    \hline
    variable & cut \\  \hline \hline
 \multicolumn{2}{|c|}{general events} \\ \hline
    $E_{\mathrm{ charged}}^{\mathrm{ hemisph.}}$ & $ \ge 0.03\cdot \sqrt{s}$  \\
      \hline
    $E_{\mathrm{ charged}}^{\mathrm{ total}}$ & $ \ge 0.12\cdot \sqrt{s}$  \\
      \hline
    $N_{\mathrm{ charged}}$ & $\ge 5$ \\
      \hline
    $\vartheta_{\mathrm{ sphericity}}$ &  $30^\circ-150^\circ$  \\
      \hline
    $p_{\mathrm{max}}$ &  $45$ GeV  \\
      \hline
 \multicolumn{2}{|c|}{three-jet events} \\ \hline
    $\sum_{i=1}^{3}\theta_i$ & $ > 355^\circ$ \\
      \hline
    $E_{\mathrm{ visible}}/\mathrm{ jet}$ & $ \ge 5$ GeV  \\
      \hline
    $N_{\text{charged}}/\mathrm{ jet}$ & $\ge 2$ \\
      \hline
    $\vartheta_{\mathrm{ jet}}$
            &  $30^\circ-150^\circ$      \\
      \hline
  \end{tabular}  
  \caption{\label{t:evtcut} Selection cuts applied to general events and 
to three-jet events.}
\end{center} 
\end{minipage}\vspace*{-.5cm}
\end{table}
The general event cuts shown in 
Table \ref{t:evtcut} select hadronic decays of the $Z$ and suppress
background from leptonic $Z$ decays, $\gamma\gamma$ interactions
or beam-gas interactions to a negligible level. Further reduction of
background is achieved by the jet-selection cuts given also in Table 
\ref{t:evtcut}.
The cut variables are the visible charged energy, 
$E^{\text{total}}_{\text{charged}}$ and $E^{\text{hemisph.}}_{\text{charged}}$,
observed in the
event or in one event hemisphere, respectively. 
Event hemispheres are defined by 
the plane perpendicular to the sphericity axis.
The polar angle of this axis with respect to the beam is 
$\vartheta_{\text{sphericity}}$ and $N_{\text{charged}}$ 
is the observed charged multiplicity.
Events are discarded, if they
contain charged particles with momenta above the kinematic limit.

Events are then clustered into jets using the aoD (or, alternatively, the
Durham or Cambridge) algorithm. 
The three-jet event quality requirements also shown in
Table \ref{t:evtcut} assure well-measured jets. 
Here $\theta_i$ denotes 
the angle between the two jets opposite to jet $i$,
$\vartheta_{\text{jet}}$ the polar angle of a jet
and  $E_{\text{visible}} /\text{jet}$ the total visible energy per jet. 
The sequence number $i$ of 
a jet is
given by ordering the jets with decreasing jet energy $(E_1>E_2>E_3)$.
These exact jet energies were calculated from the inter-jet angles assuming
massless kinematics (see e.g. \cite{scalingpaper}).
The quality of the selected two-jet events is guaranteed already by the event
cuts.

Besides general three-jet topologies, 
 mirror-symmetric three-jet topologies 
(defined by $\theta_3-\theta_2 \leq 5^{\circ}$) 
are considered in later cross-checks. 
Symmetric three-jet topologies form a class of especially simple and 
clean three-jet 
topologies which we used extensively in previous studies 
\cite{glupaper1,scalingpaper,Abreu:1999rs}.
 Therefore, events with
symmetric topologies are considered separately in order to compare the
results obtained with this clean, but statistically limited class of
events with the results obtained from events with more general topologies.
Note, that the symmetric-event sample is almost completely included
in the general sample. However, it
allows the angular range to be extended to lower values of $\theta_1$.

Initially 3,150,000 events measured by \delphi\ enter into the analysis.
In total 1,797,429 $Z$ events fulfill the above event cuts, of which
1,031,080 are accepted as two-jet events, 309,227 as general 
and 53,344 as symmetric three-jet events.

The charged-particle multiplicity  
in cones perpendicular to the event axis of a two-jet event or the
event plane of a three-jet event
has then been determined. 
Besides the cuts mentioned above it was required that the axis of 
the cones has a polar angle of at least $30^{\circ}$ with respect to the beam
in order to assure a sufficient acceptance for particle tracks.

The multiplicity has been corrected for the applied 
cuts and the limited detector acceptance and resolution by a
multiplicative correction factor. Using simulated events
this has been calculated  as
the ratio of the multiplicities of generated events
to accepted events at the detector level.
The events were generated with the \jetset\ 7.3
parton-shower model \cite{jetset} as tuned to \delphi\ data
\cite{generatorpaper}. The influence of the 
magnetic field, detector material, signal generation and
digitisation was simulated. The simulated data
were then treated like the measured events.

It has already been shown \cite{tune-incoh,generatorpaper}, 
that the Monte Carlo model underestimates the 
production of particles
at large angles with respect to the primary partons. 
Therefore, the hadron multiplicity
in cones perpendicular to the two-jet event axis or the three-jet
event plane has been reweighted with a small overall 
multiplicative correction. 
It was deduced from a comparison
of data and simulation for two-jet events. After
correction the model well describes the data and its 
dependences on the three-jet topology.

In particular, due to the 400 \mev\ momentum cut on the charged particles 
the correction factors  obtained for the charged multiplicity in the cones 
are comparably  big ($\sim 2$), but
vary only slightly with event topology.
Moreover, in the important ratio of the multiplicity of
three- and two-jet events the correction factors cancel to a large
extent. 
The correction factor for the ratio $N_3/N_2$  
is only $\sim 0.9$.
The stability of the acceptance correction has been tested by varying
the selections imposed on charged-particle tracks shown in 
Table \ref{t:tcuts}. 
In particular the lower momentum cut has been changed by $\pm 100 \text{ MeV}$.
No significant change in the results was observed.
Conservatively a 10\% relative uncertainty of the respective 
acceptance correction
is assumed to propagate to the quoted multiplicities or multiplicity ratios.
As this uncertainty is expected to mainly change the normalisation of the 
multiplicities it is
not included in the overall uncertainties shown in the figures.
It is, however, considered in the results obtained from
global fits of the data.

\section{Results \label{results}}
Experimental results on the average charged-hadron multiplicities in cones 
are presented for 
cones perpendicular to the event axis in two-jet events and cones perpendicular
to the event plane in three-jet events. Moreover the ratio of the cone
multiplicities in three- and two-jet events is given as well as
the momentum spectra in cones perpendicular to the plane of three-jet events.

The charged-hadron multiplicity in cones perpendicular to the  event
axis in two-jet events, $N_2$, is given in Table \ref{t:2jet}
for three different cone opening angles and three different 
values  of the jet-resolution parameter $y_{\text{cut}}^{\text{aoD}}$.
The multiplicity shows the expected slight increase with $y_{\text{cut}}$ 
and, due to the 
increase of phase-space, an approximately quadratic 
increase with cone opening angle.

 \begin{table}[b]
\newcommand{\haho}[1]{\raisebox{1.5ex}[-1.5ex]{#1}}
 \begin{center}
 \begin{tabular}{|c||c|c|c|}
 \hline
 & \multicolumn{3}{|c|}{$y_{\text{cut}}$} \\ \cline{2-4}
 \haho{$\vartheta_{cone}$}
 & 0.01
 & 0.015
 & 0.02
 \\ \hline \hline
 {$20^\circ$}& $0.231\pm  0.001$
 & $0.245\pm  0.001$
 & $0.255\pm  0.001$
 \\ \hline
 {$30^\circ$}& $0.537\pm  0.001$
 & $0.570\pm  0.001$
 & $0.593\pm  0.001$
 \\ \hline
 {$40^\circ$}& $1.007\pm  0.002$
 & $1.067\pm  0.002$
 & $1.111\pm  0.002$
 \\ \hline
 \end{tabular}
 \end{center}
\caption
{\label{t:2jet} 
Charged-hadron multiplicity in cones perpendicular 
to the  event axis in two-jet events, $N_2$,
as a function of the cone opening angle and the jet-resolution
for the aoD algorithm. Errors are statistical.} 
\end{table}

\begin{figure}[tbhp]
\unitlength1cm
 \unitlength1cm
 \begin{center}
           \mbox{\epsfig{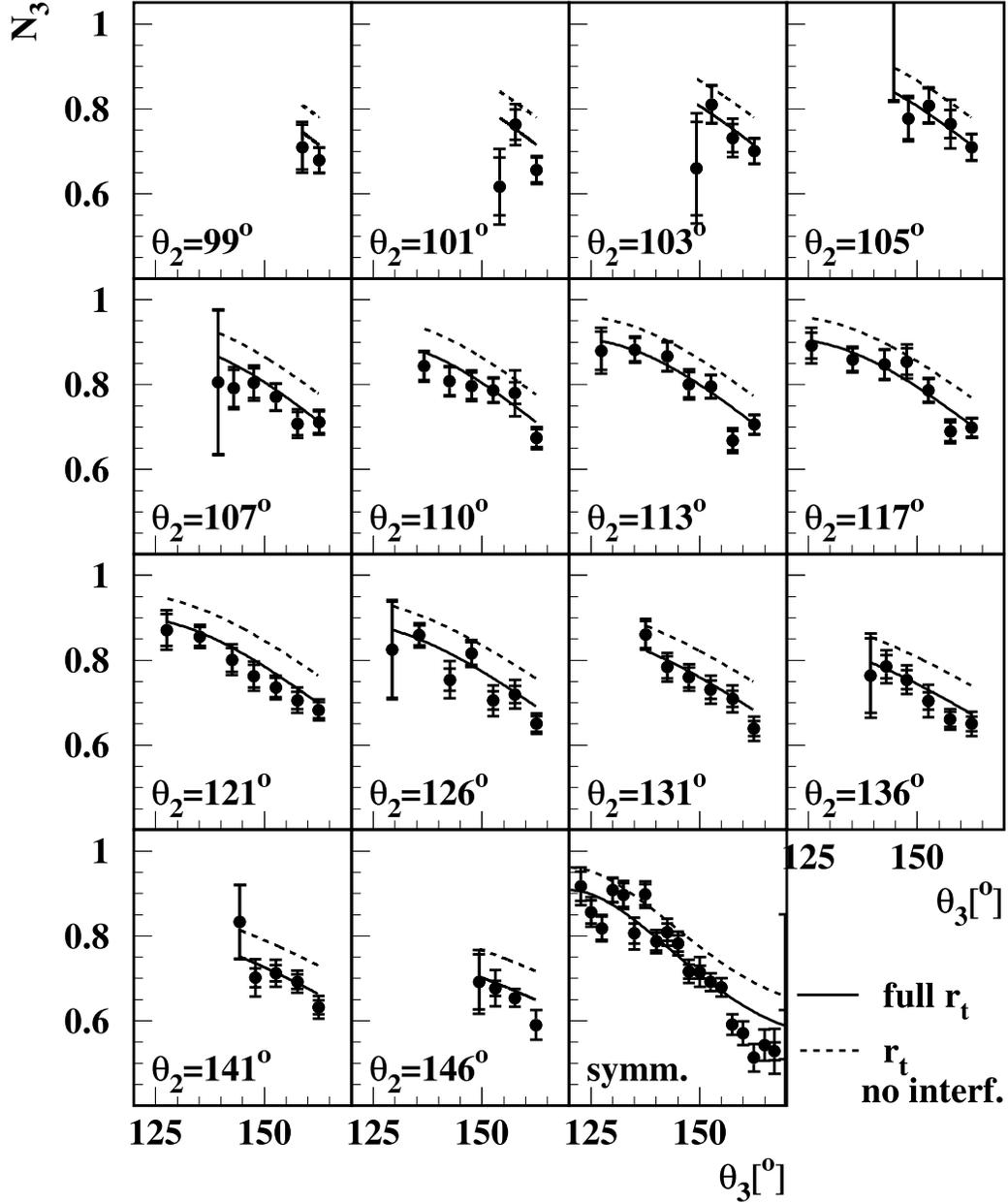}}
\end{center}
\caption{\label{f:cone_pred}
Multiplicity in cones of $30^{\circ}$ opening angle perpendicular to the 
three-jet event plane, $N_3$, as a function of the 
opening angles $\theta_2, \theta_3$.
The inner error bars are statistical, the outer also include systematic
uncertainties (see text). The full line is the expectation deduced
from Equation {\protect \ref{e:basic}} using $N_2$ from Table \ref{t:2jet}.
For the dashed line the interference term in Equation \ref{e:basic}
($\propto 1/N_c^2$) has been omitted.
}
\end{figure}

The corresponding three-jet multiplicity, $N_3$, is
shown in Figure \ref{f:cone_pred} as a
function of the inter-jet angles $\theta_2$ and $\theta_3$
for general three-jet topologies and as a function of $\theta_3$ for 
symmetric ones. The cone opening angle is
$\vartheta_{\text{cone}}=30^{\circ}$ and $y_{\text{cut}}^{\text{aoD}}=0.015$. 
For the general three-jet events
the bin width in $\theta_2$ is: 
2\degs\ for $98^\circ <\theta_2 <108^\circ$,
3\degs\ for $108^\circ <\theta_2 <111^\circ$,
4\degs\ for $111^\circ <\theta_2 <123^\circ$ and
5\degs\ for $123^\circ <\theta_2 <148^\circ$.
The corresponding binning in $\theta_3$ is:
10\degs\ for $120^\circ <\theta_3 < 140^\circ$ and
5\degs\ for $140^\circ <\theta_3 <165^\circ$.
For symmetric topologies the 
range in $\theta_1$ is fixed by the 
highest accessible
angle of  120\degs\ and the bin width of 5\degs\ imposed by the condition 
$\theta_3-\theta_2 \leq 5^{\circ}$. The lower limit of $\theta_1\sim$20\degs\ 
depends on the applied value of $y_{cut}$.
Numerical values of the results are given in \cite{martin}.
The inner error bars 
represent the statistical error, the outer bars also include the systematic
uncertainty relevant for the comparison to the theoretical expectation of
Equation \ref{e:basic}, added in quadrature. 
As only the ratio of multiplicities in three- and two-jet events enters in 
the comparison to the theoretical expectation (see Equation \ref{e:basic}),
the relative systematic 
uncertainty of the multiplicity has been taken to be identical to the
relative error of the three- to two-jet multiplicity ratio.
It has been  determined as the r.m.s. of the results obtained when 
varying the cone size, jet-resolution and jet algorithm as described in
Section \ref{theo}. 

The theoretical expectation, shown as a full line in Figure \ref{f:cone_pred},
describes the data well for all topologies except for large $\theta_3$ 
for symmetric three-jet topologies. This
region corresponds to small $\theta_1$, i.e. close-by low-energy jets.
In this case the event plane is ill determined and
data and theory cannot be compared reliably.

The prediction increases (by $\sim 7\%$, see dashed line in 
Figure \ref{f:cone_pred}) if the interference term 
of the theoretical prediction $\propto 1/N_c^2$ is omitted.
In this case the prediction is significantly above the data.

In order to quantify this result the interference term in Equation
\ref{e:basic} has been multiplied by an amplitude factor $k$ and then fitted
to the ratio 
$N_3/N_2$ of the multiplicities as given in 
Figure \ref{f:cone_pred} and Table \ref{t:2jet}, respectively.
The fit yields:
\begin{xalignat*}{3}
k&= 1.37 \pm 0.05 {\text{ (stat.)}} \pm 0.33{\text{ (syst.)}} 
& \chi^2/Ndf &= 1.2\\
\intertext{for general three-jet topologies and}
k&= 1.23 \pm 0.13 {\text{ (stat.)}} \pm 0.32 {\text{ (syst.)}} 
& \chi^2/Ndf &= 2.1 
\end{xalignat*}
for symmetric three-jet events with $\theta_1 > 48^{\circ}$.
The systematic uncertainty of $k$ was taken to be  
the r.m.s. of the results obtained when 
varying the cone size, jet-resolution and jet algorithm as described in
Section \ref{theo}. 
Moreover a 10\% uncertainty on the acceptance correction applied for the
multiplicity ratio has been assumed.
The $\chi^2$ has been calculated using statistical errors only.
Therefore, a significant destructive interference term is observed in the 
data-to-theory comparison. 
The size of the term is consistent within error with 
expectation. 
Moreover it has been verified that the topology dependence of the 
$1/N_c^2$ term is
consistent with that present in the data.
This result represents 
the first direct verification of a colour suppressed negative
antenna contribution not included in the usual 
probabilistic parton cascade calculation.
Note, that it is based on the direct comparison of 
data to an absolute
prediction. In particular no construction of an unphysical incoherent model
is needed for the comparison.

\begin{figure}[bt]
\unitlength1cm
 \unitlength1cm
 \begin{center}
           \mbox{\epsfig{file=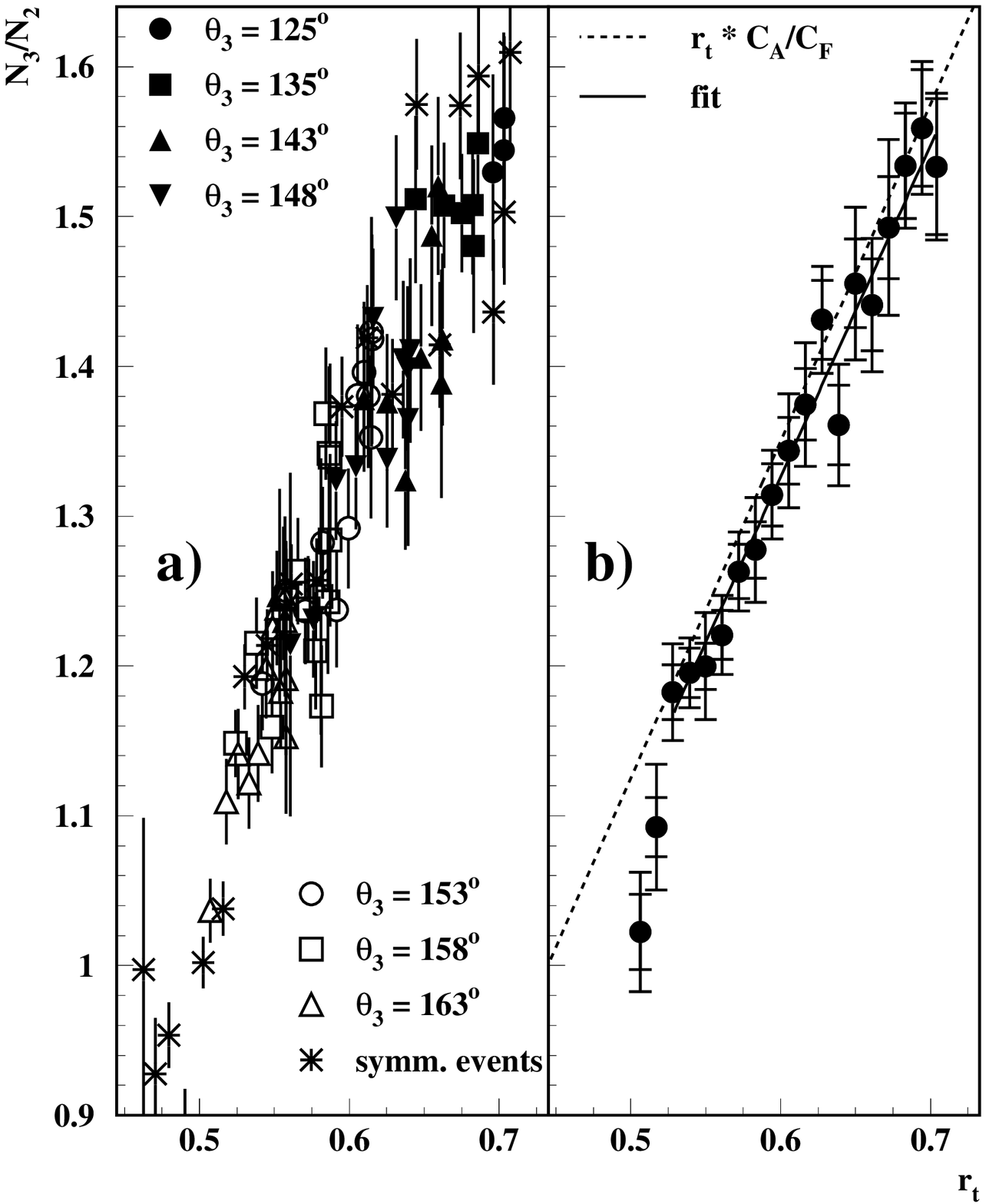,width=10cm}}
\end{center}
\caption{\label{f:cone_line}
Multiplicity ratio $N_3/N_2$ in cones of $30^{\circ}$ opening angle 
as function of $r_t$:
a) for different angles $\theta_3$, errors are statistical;
b) averaged over $\theta_3$, inner errors are statistical, outer
errors include systematic uncertainties (see text).
The dashed line is the expectation Equation {\protect \ref{e:basic}},
the full line is a fit to the data in the indicated $r_t$-interval.
}
\end{figure}

Figure \ref{f:cone_line} a) shows the multiplicity ratio $N_3/N_2$ 
as function of $r_t$ for different $\theta_3$.
The data for different values of $\theta_3$ agree well as is already expected
from the good description of the data by the expectation, Equation
\ref{e:basic}, shown in Figure \ref{f:cone_pred}. 
The exclusive dependence on $r_t$ as determined from the general-event 
sample is  shown in 
Figure \ref{f:cone_line} b) compared to the expectation, i.e. to the
homogeneous straight-line $N_3/N_2=C_A/C_F\cdot r_t$ indicated dashed.
The deviation of the points at low values of 
$r_t$  is again due to events with close-by jets.
A homogeneous straight line fit to the data for 
$r_t > 0.525$ yields for the slope:
\begin{xalignat*}{3}
&2.211 \pm 0.014 \text{(stat.)} \pm 0.053 \text{(syst.)} & \chi^2/Ndf &=1.3 
\end{xalignat*}
in astonishingly good agreement with the 
leading-order QCD expectation \cacf$=2.25$. 
The systematic uncertainty was again determined as for $k$ in 
the interference term.
The fit result is indicated by
the full line in Figure \ref{f:cone_line} b).
This result for the first time quantitatively 
verifies the colour-factor ratio \cacf\   
from hadronic multiplicities and a leading-order prediction.
This simple prediction applies, because higher order contributions
prove to be unimportant for soft particles emitted at large
angles \cite{ochskhoze}. 
This is different from the more global case which depends on
all the radiation inside jets.
Moreover, these particles are least
affected by energy-conservation and leading-particle effects.

For completeness it should be mentioned that the data do not allow 
the colour-factor ratio and the coherence term to be simultaneously 
verified
due to the obvious high correlation of both terms.
If such a fit is attempted the data can as well be described with
\cacf$\sim 2 $ and $N_C\rightarrow \infty$. 
This result is also expected 
from gauge theories with a large number of colours. 
It is clear, however, that this solution does not apply as $N_C=3$
is well known experimentally, e.g. from the ratio, $R$, of the hadronic to
the leptonic cross-section at the $Z$ pole.

\begin{figure}[t]
\unitlength1cm
 \unitlength1cm
 \begin{center}
           \mbox{\epsfig{file=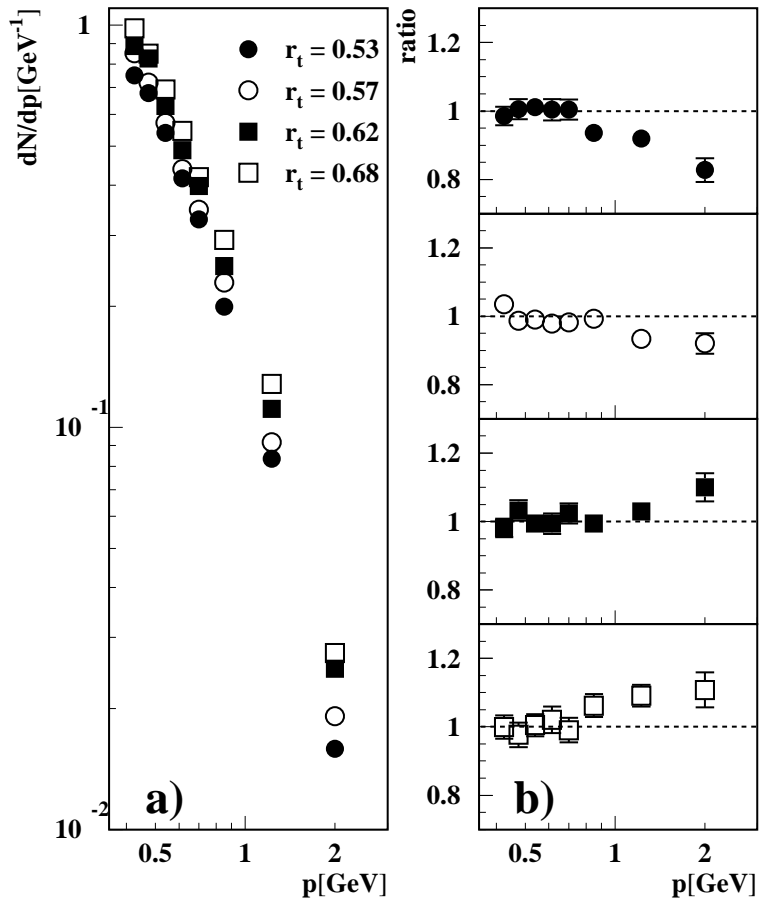,width=10cm}}
\end{center}
\caption{\label{f:pdis}
a) Differential momentum distribution of hadrons 
in cones of $30^{\circ}$ opening angle perpendicular to the 
three-jet event plane for different values of
$r_t$ (aoD algorithm, $y_{cut}=0.015$). 
b) Ratio of the momentum distributions to their average
reweighted using Equation \ref{e:basic} and the individual $r_t$ values.
}
\end{figure}

The dependence of the multiplicity on the scale $r_t$ only allows
inclusive distributions of the produced hadrons to be studied.
In Figure \ref{f:pdis} a) the differential momentum distribution is shown
for different values of $r_t$. The momentum distributions are observed to
scale approximately
with $r_t$. This is most clearly seen in Figure \ref{f:pdis} b)
where the ratio of the momentum spectra to the average 
spectrum is shown.
Moreover, these ratios have been scaled by $\langle r_t \rangle/r_t$ such that 
a unit value for the overall multiplicity ratio is expected.
For small momenta ($p < 1$~GeV) the data are consistent with unity, for higher
momenta less (more) particles are produced for small (high) scales $r_t$,
respectively. 
The constancy of the ratio is expected if the production of hadrons is
directly proportional to gluon radiation, thus the observed result 
strongly supports the 
conjecture of local parton-hadron duality, LPHD \cite{lphd},
for low-energy hadrons. The deviation from unity for higher momenta
may be understood as being due to the increase of the 
hadronic phase-space for high $r_t$.

In summary the multiplicity in cones perpendicular to the event plane of
three-jet events produced in \epem\ annihilation has been measured for the
first time and compared to the corresponding multiplicity in two-jet events.
The multiplicity ratio is well described by a leading-order QCD prediction
also including a destructive interference term. The presence of destructive 
interference and thereby of the coherent nature of soft hadron production 
was verified from the comparison of the data to the 
absolute QCD prediction. 
The slope of the hadron multiplicity with the topological variable $r_t$
directly represents the colour-factor ratio \cacf. This measurement 
for the first time is able to verify the colour-factor ratio using a 
hadron multiplicity measurement and a leading-order QCD prediction. 
This possibility is a consequence of the favorable perturbative situation 
studied and
the focus on low-energy large-angle particles.
A simultaneous study of the momentum and topology  dependence of the hadron 
multiplicity provides evidence for the LPHD
conjecture.

%
\subsection*{Acknowledgements}
\vskip 3 mm
 We are greatly indebted to our technical 
collaborators, to the members of the CERN-SL Division for the excellent 
performance of the LEP collider, and to the funding agencies for their
support in building and operating the DELPHI detector.\\
We acknowledge in particular the support of \\
Austrian Federal Ministry of Education, Science and Culture,
GZ 616.364/2-III/2a/98, \\
FNRS--FWO, Flanders Institute to encourage scientific and technological 
research in the industry (IWT), Belgium,  \\
FINEP, CNPq, CAPES, FUJB and FAPERJ, Brazil, \\
Czech Ministry of Industry and Trade, GA CR 202/99/1362,\\
Commission of the European Communities (DG XII), \\
Direction des Sciences de la Mati$\grave{\mbox{\rm e}}$re, CEA, France, \\
Bundesministerium f$\ddot{\mbox{\rm u}}$r Bildung, Wissenschaft, Forschung 
und Technologie, Germany,\\
General Secretariat for Research and Technology, Greece, \\
National Science Foundation (NWO) and Foundation for Research on Matter (FOM),
The Netherlands, \\
Norwegian Research Council,  \\
State Committee for Scientific Research, Poland, SPUB-M/CERN/PO3/DZ296/2000,
SPUB-M/CERN/PO3/DZ297/2000, 2P03B 104 19 and 2P03B 69 23(2002-2004)\\
FCT - Funda\c{c}\~ao para a Ci\^encia e Tecnologia, Portugal, \\
Vedecka grantova agentura MS SR, Slovakia, Nr. 95/5195/134, \\
Ministry of Science and Technology of the Republic of Slovenia, \\
CICYT, Spain, AEN99-0950 and AEN99-0761,  \\
The Swedish Natural Science Research Council,      \\
Particle Physics and Astronomy Research Council, UK, \\
Department of Energy, USA, DE-FG02-01ER41155. \\
EEC RTN contract HPRN-CT-00292-2002.\\
We thank V. Khoze and W. Ochs for
numerous enlightening discussions and helpful explanations and
comments.


\end{document}